\documentclass[twocolumn,pra,aps,showpacs]{revtex4-1}
\usepackage{amsfonts}
\usepackage{amssymb}
\usepackage{epsfig}
\usepackage{amsbsy}
\usepackage{amssymb}
\usepackage{amsmath}
\usepackage{graphicx}
\usepackage{graphics}

\newcommand{\stensor}{\overleftrightarrow}
\newcommand{\ie}{\textit{i.e. }}
\newcommand{\sem}{\text{(em)}}
\newcommand{\md}{\mathrm{d}}
\newcommand{\etal}{\textit{et al.}}

\begin{document}

\title{Theory of Optomechanical Interactions in Superfluid He}
\author{G. S. Agarwal}
\affiliation{Department of Physics, Oklahoma State University, Stillwater, Oklahoma 74078, USA}
\author{Sudhanshu S. Jha}
\affiliation{UM-DAE Centre for Excellence in Basic Sciences, University of Mumbai Vidyanagari Campus, Mumbai 400098, India}
\date{\today }

\begin{abstract}
  A general theory is presented to describe optomechanical interactions of acoustic phonons, having extremely long lifetimes in superfluid $^4$He, with optical photons in the medium placed in a suitable electromagnetic cavity. The acoustic nonlinearity in the fluid motion is included to consider processes beyond the usual linear process involving absorption or emission of one phonon at a time. We first apply our formulation to the simplest one-phonon process involving the usual resonant anti-Stokes upconversion of an incident optical mode. However, when the allowed optical cavity modes are such that there is no single-phonon mode in the superfluid which can give rise to a resonant allowed anti-Stokes mode, we must consider the possibility of  two-phonon upconversion. For such a case, we show that the two-step two-phonon process could be dominant. We present arguments for large two-step process and negligible single step two-phonon contribution. The two-step process also shows interesting quantum interference among different transition pathways.
\end{abstract}
\pacs{42.50.Wk, 07.10.Cm, 42.65-k.}
\maketitle


\section{Introduction}
In the field of optomechanics, one is always designing mechanical systems with lowest possible friction~\cite{PNAS,Barker} and highest possible optomechanical coupling constant~\cite{Aspelmeyer,Harris,Vitali}. This is because one would like to produce and use coherent phonons with long coherence time. One prominent application of coherent phonons is in storage and retrieval of light using optomechanical systems~\cite{EIT,storage}.  An attractive system which has received considerable attention is the levitated microsphere trapped in an optical cavity. Both trapping and levitation can be produced by optical fields~\cite{PNAS,Barker}. Another very attractive system is superfluid He, which has zero viscosity. It is known to have acoustic phonons with almost zero friction at low temperatures, a finite value arising only from thermal three-phonon scattering processes~\cite{Schwab}. De Lorenzo and Schwab~\cite{Schwab} have performed initial optomechanical experiments on superfluid He by coupling it to a superconducting resonator. Flowers-Jacobs \etal~\cite{meeting} have reported progress in doing optomechanics with superfluid He using optical cavities.

In view of the current interest~\cite{Schwab,meeting} in the optomechanics with superfluid He, we present in this paper theoretical foundations of optomechanics in such systems. The organization of this paper is as follows. In Sec. II, we derive the basic semiclassical equations for the optomechanical interactions in superfluid He. In Sec. III, we present a Hamiltonian formulation of the problem in terms of the canonical variables, so that this can be adopted for quantized phonon and photon fields. The theory is formulated in terms of the fields, both electromagnetic and fluid density, so that situations involving many phonons and photons of different frequencies can be handled. In Sec. IV, we present a quantized description of the optomechanical interactions. We present estimates for the strength of the optomechanical interactions. The linear optomechanical interaction---shift of the cavity resonance per photon---is quite significant in cavities like a fiber cavity~\cite{Jacobs}. We derive the canonical form~\cite{OMS1,OMS2,OMS3,OMS4} of the Hamiltonian for linear optomechanical interactions in superfluid He. Having obtained the canonical form, we can study all the physical processes that have been studied with other optomechanical systems. An estimate of the single step two-phonon antiStokes process due to acoustic nonlinearity is also given in this section. In Sec. V, we discuss two-step two-phonon processes, which are shown to be significant in superfluid He. It may be noted that the knowledge of the linear interaction Hamiltonian (\ref{04.19}) with the strength of $g$ estimated after Eq (\ref{04.15}) is sufficient to understand the processes in Sec. V. When the electromagnetic cavity is designed in such a way that allowed optical modes are such that no anti-Stokes upconversion is possible via the absorption of any single phonon in the medium, one must consider absorption of two phonons for possible upconversion. Such a process can be controlled well when these phonons are external phonons injected in the medium.  Because of the intrinsic nonlinearity of the superfluid He, we have the new possibilities arising from the combination of the superfluid nonlinearity and the optomechanical interactions.

\section{Classical Nonlinear Equations for Superfluid Helium Optomechanics}
In this section, we start with the fundamental equations~\cite{London} for the superfluid density $\rho$ and the velocity $\vec{v}$ and we obtain modifications of these due to interaction with the electromagnetic fields. The basic equations for $\rho$ and $\vec{v}$ in the absence of the electromagnetic fields are given by
\begin{align}
  \frac{\partial\rho}{\partial t} + \vec{\nabla}\cdot(\rho\vec{v}) &= 0, \label{2.1}
\end{align}
\begin{align}
  \frac{\partial}{\partial t}(\rho\vec{v}) + \vec{\nabla}\cdot \stensor{T} &= 0, \label{2.2}
\end{align}
where the stress tensor $\stensor{T}$ is given by
\begin{equation}\label{2.3}
  T_{ij} = p\delta_{ij} + \rho v_i v_j.
\end{equation}
Here $p$ is the pressure in the superfluid. We have set the viscosity term zero. We note that we have a set of nonlinear equations as pressure is generally expanded~\cite{Wright,Abraham} in terms of the normalized deviation $\tilde{\rho}=(\rho-\rho_0)/\rho_0$ from the equilibrium value $\rho_0$:
\begin{equation}\label{2.4}
  p - p_0 \approx \left(\rho\frac{\partial p}{\partial\rho}\right)_0 \tilde{\rho} + \frac12 \left(\rho^2\frac{\partial^2 p}{\partial\rho^2}\right)_0 \tilde{\rho}^2 + \cdots.
\end{equation}
We now discuss the modification of Eqs.~(\ref{2.1}) and (\ref{2.2}) due to the interaction with the electromagnetic fields. Clearly Eq.~(\ref{2.1}) remains unchanged. We need to modify Eq.~(\ref{2.2}) by the addition of the Maxwell stress contribution $\stensor{T}^\sem$ to Eq.~(\ref{2.2}) \ie
\begin{equation}\label{2.5}
  \frac{\partial}{\partial t}(\rho\vec{v}) + \vec{\nabla}\cdot \stensor{T} - \vec{\nabla}\cdot \stensor{T}^\sem = 0.
\end{equation}
The form of the Maxwell stress tensor depends on the nature of the medium. It is derived from the considerations of the electromagnetic force on the medium. On dropping the magnetic polarization contribution, the force on a linear medium can be written as
\begin{equation}\label{2.6}
  \begin{aligned}
    F_j^\sem &= \int \md ^3r \sum_i \frac{\partial}{\partial r_i} T^\sem_{ij}, \\
    T^\sem_{ij} &= -\frac12 \delta_{ij}\epsilon(\vec{r})\vec{E}\cdot\vec{E} + \epsilon(\vec{r})E_iE_j.
  \end{aligned}
\end{equation}
Here $\vec{E}$ is the electromagnetic field and $\epsilon(\vec{r})$ is the optical dielectric function of the isotropic superfluid. The electromagnetic force can also be written in an alternate form~\cite{Stratton}
\begin{equation}\label{2.7}
  \vec{F}^\sem = -\frac12\int \md ^3r E^2(\vec{r}) \vec\nabla\epsilon(\vec{r}).
\end{equation}
The dielectric function of the medium depends on $\vec{r}$ through the density \ie $\epsilon(\vec{r}) = \epsilon[\rho(\vec{r})]$ and hence
\begin{equation}\label{2.8}
  \vec\nabla\epsilon(\vec{r}) = \frac{\partial \epsilon[\rho(\vec{r})]}{\partial\rho} \vec\nabla\rho,
\end{equation}
and hence Eq.~(\ref{2.7}) reduces to
\begin{align}\label{2.9}
  \vec{F}^\sem &= -\frac12\int \md ^3r E^2(\vec{r}) \frac{\partial \epsilon}{\partial\rho} \vec\nabla\rho \nonumber \\
  &= \frac12\int\md ^3r \rho(\vec{r}) \vec\nabla \left[\frac{\partial \epsilon}{\partial\rho}E^2(\vec{r})\right].
\end{align}
Using Eq.~(\ref{2.9}), Eq.(\ref{2.5}) becomes
\begin{equation}\label{2.10}
  \frac{\partial}{\partial t}(\rho\vec{v}) + \vec{\nabla}\cdot \stensor{T} + \frac12\rho\vec{\nabla}\left[\frac{\partial \epsilon}{\partial\rho}E^2(\vec{r})\right] = 0.
\end{equation}
The equations Eqs.~(\ref{2.1}) and (\ref{2.10}) are the basic equations for the optomechanical interactions in superfluid He. The only assumption that we made in deriving Eq.~(\ref{2.10}) is the linear electromagnetic response $\epsilon[\rho(\vec{r})]$ of superfluid He. The equations (\ref{2.1}) and (\ref{2.10}) are to be supplemented by the expansion (\ref{2.4}). The electric field obeys the equation
\begin{equation}\label{2.11}
  \vec{\nabla}\times\vec{\nabla}\times\vec{E} + \frac{1}{\epsilon_0c^2}\frac{\partial^2}{\partial t^2}\left(\epsilon[\rho(\vec{r})]\vec{E}\right) = 0,
\end{equation}
which is obtained from the Maxwell equations.

\section{The Hamiltonian description of the basic Eqs. (\ref{2.10}) and (\ref{2.11})}
In the Hamiltonian description, one introduces the conjugate variables $\rho(\vec{r})$ and $\Phi(\vec{r})$ and the classical velocity is related to $\Phi$ via
\begin{equation}\label{3.1}
  \vec{v} = -\vec\nabla\Phi.
\end{equation}
The Hamiltonian description of the superfluid equations (\ref{2.1}) and (\ref{2.2}) is well-known and for completeness, we recall the main aspects. The unperturbed Hamiltonian density is
\begin{equation}\label{3.2}
  \mathcal{H}_0 = \frac12 \rho(\nabla\Phi)^2
\end{equation}
and the interaction term is
\begin{equation}\label{3.3}
  \mathcal{H}_1 = \rho W(\rho).
\end{equation}
The function $W$ is related to the pressure via the thermodynamic relation~\cite{Abraham}
\begin{equation}\label{3.4}
  W(\rho) = \int_{\rho_0}^\rho \frac{p(\rho')}{\rho'^2} \md \rho'.
\end{equation}
Using the total Hamiltonian $H=\int\md ^3r (\mathcal{H}_0+\mathcal{H}_1)$, we can see how Eqs.(\ref{3.2})-(\ref{3.4}) lead to Eqs.(\ref{2.1}) and (\ref{2.2}). For this purpose, we use the Hamiltonian formulation for fields~\cite{Goldstein}
\begin{align}
  \dot{\rho} &= -\frac{\delta H}{\delta\Phi} = \sum_j\frac{\partial}{\partial r_j} \left(\frac{\partial H}{\partial(\partial\Phi/\partial r_j)}\right) - \frac{\partial H}{\partial\Phi} \nonumber\\
  &= \sum_j\frac{\partial}{\partial r_j} \left[ \rho\frac{\partial\Phi}{\partial r_j}\right] = -\sum_j\frac{\partial}{\partial r_j} [\rho v_j] \nonumber\\
  &= -\vec{\nabla}\cdot(\rho\vec{v}), \label{3.5} \\
  \dot{\Phi} &= \frac{\delta H}{\delta\rho} = \frac12(\nabla\Phi)^2 + \frac{\partial}{\partial\rho} [\rho W(\rho)] \nonumber\\
  &= \frac12v^2 + \left(W + \frac{p(\rho)}{\rho} \right). \label{3.6}
\end{align}
We can convert Eq.(\ref{3.6}) into an equation for $(\rho\vec{v})$ as follows
\begin{align}\label{3.7}
  \frac{\partial}{\partial t}(\rho\vec{v}) &= \frac{\partial\rho}{\partial t}\vec{v} - \rho\frac{\partial}{\partial t}\vec{\nabla}\Phi \nonumber \\
  &= -[\vec{\nabla}\cdot(\rho\vec{v})]\vec{v} - \rho\vec{\nabla}\left[\frac12v^2 + W + \frac{p(\rho)}{\rho}\right] \nonumber \\
  &= -\rho(\vec{\nabla}\rho)\frac{\partial}{\partial\rho} \left[W+\frac{p(\rho)}{\rho}\right] - \vec{\nabla}\cdot[\rho\vec{v}\vec{v}] \nonumber \\
  &= - \vec{\nabla}\cdot[\rho\vec{v}\vec{v}] - \vec{\nabla}\cdot(p\stensor{I}),
\end{align}
where $\stensor{I}$ is the unit tensor. The Eqs.(\ref{3.5}) and (\ref{3.7}) are identical to Eqs.(\ref{2.1}) and (\ref{2.2}) respectively.

The Hamiltonian for optomechanical interactions in superfluid He will then be
\begin{align}
  & H = \int \md^3r(\mathcal{H}_0+\mathcal{H}_1) + \int\mathcal{H}^\sem \md^3r, \label{3.8}\\
  &\mathcal{H}^\sem = \mathcal{H}^\sem_0 + \mathcal{H}^\sem_1,\label{3.9} \\
  &\mathcal{H}^\sem_0 = \frac12 (\epsilon_0E^2 + \frac{1}{\mu_0}B^2), \label{3.10}\\
  &\mathcal{H}^\sem_1 = -\frac12 \vec{P}\cdot\vec{E} = -(\frac{\epsilon[\rho]-\epsilon_0}{2})E^2,  \label{3.11}
\end{align}
where $\vec{P}$ is the polarization in superfluid medium. Using Eq.~(\ref{3.8}), the equations for the canonical conjugate variables $\rho$ and $\Phi$ are
\begin{align}
  \dot{\rho} - \vec{\nabla}\cdot(\rho\vec{\nabla}\Phi) =0, \label{3.12}\\
  \dot{\Phi} = \frac12(\vec{\nabla}\Phi)^2 + \frac{\partial}{\partial\rho}(\rho W) - \frac12(\frac{\partial\epsilon}{\partial\rho})E^2. \label{3.13}
\end{align}
A simple exercise shows that Eq.(\ref{3.13}) is equivalent to the Eq.(\ref{2.10}) for $\rho\vec{v}$.

The Hamiltonian (\ref{3.8}) depends on the density $\rho$ to all orders. In order to bring out some of the important physical process, we consider an expansion of $H$ in powers of the deviation $\tilde\rho$, $(\rho-\rho_0)/\rho_0$ from the equilibrium value $\rho_0$. We will examine terms up to second order in $\tilde\rho$. The expansion of the optomechanical interaction term is straight forward:
\begin{gather}
  \epsilon[\rho] = \epsilon[\rho_0] + \rho_0\left(\frac{\partial\epsilon}{\partial\rho}\right)_0\tilde\rho + \frac12 \rho_0^2\left(\frac{\partial^2\epsilon}{\partial\rho^2}\right)_0\tilde{\rho}^2 + \dots, \label{3.14}\\
  \mathcal{H}^\sem_1 = - \frac{\epsilon(\rho_0)-\epsilon_0}{2}E^2 - \frac12 g_1\epsilon_0 \tilde\rho E^2 - \frac12 g_2\epsilon_0 \tilde\rho^2 E^2  + \dots, \label{3.15} \\
  g_1 = \frac{\rho_0}{\epsilon_0}\left(\frac{\partial\epsilon}{\partial\rho}\right)_0, \qquad g_2 = \frac{\rho_0^2}{2\epsilon_0}\left(\frac{\partial^2\epsilon}{\partial\rho^2}\right)_0. \nonumber
\end{gather}
Here $g_1$ and $g_2$ are the coupling constants for the linear and quadratic optomechanical interactions. A rough estimate of $g_1$ and $g_2$ can be obtained from the experimental data~\cite{Meyer} on liquid He:
\begin{equation}\label{3.16}
  \frac{\epsilon(\rho)}{\epsilon_0} = \frac{1+\frac{8\pi}{3}\frac{\alpha_m}{m}\rho}{1-\frac{4\pi}{3}\frac{\alpha_m}{m}\rho},
\end{equation}
where the molecular polarizability $\alpha_m$ is ${1.23296\times10^{-7}}$m$^3/$mole, $m=4.0026\times10^{-3}$kg/mole, equilibrium density $\rho_0=145.1397$kg/m$^3$, and hence
\begin{equation}\label{3.17}
  g_1 \cong 0.05826, \qquad g_2 \cong 0.00111.
\end{equation}
Further, $\epsilon(\rho_0)/\epsilon_0=1.057$ and therefore the term $-\frac{\epsilon(\rho_0)-\epsilon_0}{2}E^2$ contributes to small frequency shifts of the electromagnetic fields. We will ignore such frequency shifts. The term $\mathcal{H}_1$ gives the nonlinearities of the superfluid in the absence of any applied electromagnetic fields. We write $p(\rho)$ as
\begin{align}\label{3.18}
  p(\rho) &= \rho_0\frac{\partial p}{\partial\rho_0}\tilde\rho + \frac12 \rho_0^2\frac{\partial^2 p}{\partial\rho_0^2}\tilde\rho^2 + \dots \nonumber \\
  &= (\rho_0v_s^2)\tilde\rho + \frac12 A_2\tilde{\rho}^2 + \dots,
\end{align}
and use $W(\rho_0)=0$, to obtain
\begin{equation}\label{3.19}
  \mathcal{H}_1 = \frac12(\rho_0v_s^2)\tilde{\rho}^2 + \frac16(A_2-\rho_0v_s^2)\tilde{\rho}^3 + \dots.
\end{equation}
The parameter $A_2/(2\rho_0v_s^2)$ is called the Gruneisen constant~\cite{Abraham} and has the value $2.84$. In Eq.(\ref{3.18}), ${v_s(=238}$m/sec) is the velocity of sound. The nonlinear conversion of phonons, as determined by the $\tilde\rho^3$ term in Eq.~(\ref{3.19}), has been discussed by Wright \etal~\cite{Wright}. In order to simplify $\mathcal{H}_0$ in powers of $\tilde\rho$, we need to find the expansion of $\vec\nabla\Phi$, which can be obtained from Eq.(\ref{3.12}), which to lowest order in density yields
\begin{equation}\label{3.20}
  \nabla^2\Phi = \dot{\tilde\rho}.
\end{equation}

\section{Quantization of the Hamiltonian for optomechanical interactions}
In order to do the quantization, we invoke the space-time structure of the electromagnetic and density (acoustic) fields. We would be studying optomechanical interactions in a cavity which could be an optical one like a fiber cavity~\cite{Jacobs} or a superconducting one~\cite{Schwab}. The electromagnetic field can be written as a superposition of orthogonal and orthonormal transverse modes $\vec{u}^{(i)}$ \ie
\begin{equation}\label{04.1}
  \vec E(\vec r,t) = \sum_i \vec{u}^{(i)}(\vec r) \mathcal{E}^{(i)} \mathrm{e}^{-i\omega_it} + c.c.,
\end{equation}
where the mode function $\vec{u}^{(i)}$ has frequency $\Omega_i$ and is a solution of $\nabla^2\vec{u}^{(i)} + (\Omega_i^2/c^2)\vec{u}^{(i)}=0$. The Hamiltonian (\ref{3.10}) for the electromagnetic field leads to
\begin{equation}\label{04.2}
  \mathcal{H}^\sem_0 = 2\epsilon_0\sum_i |\mathcal{E}^{(i)}|^2,
\end{equation}
where we used the orthogonality of the mode functions $\displaystyle \int[u^{(i)}(\vec r)\cdot u^{(j)*}(\vec r)]\md^3r = \delta_{ij}$. In order to do the quantization, we identify $2\epsilon_0|\mathcal{E}|^2$ with $\hbar\omega a^\dag a$. Thus the amplitude $\mathcal{E}$ is to be replaced by the annihilation operator $a$ via
\begin{equation}\label{04.3}
  \mathcal{E} \to \sqrt{\frac{\hbar\omega}{2\epsilon_0}}a.
\end{equation}
Therefore, the quantized form of the electric field is
\begin{equation}\label{04.4}
  \vec E = \sum \sqrt{\frac{\hbar\omega_i}{2\epsilon_0}} \vec{u}^{(i)}(\vec r) a_i \mathrm{e}^{-i\omega_it} + c.c.,
\end{equation}
and the unperturbed Hamiltonian is
\begin{equation}\label{04.5}
  H^\sem_0 = \sum_i\hbar\omega_i a_i^\dag a_i.
\end{equation}
We expand the phonon field in terms of the normalized mode functions $\psi_i$ with frequency $f_i$,
\begin{gather}
  \tilde\rho = \sum\psi_i(\vec r) \mathrm{e}^{-if_it} \sigma_i + c.c., \label{04.6} \\
  \nabla^2\psi_i + (f_i^2/v_s^2)\psi_i = 0. \label{04.7}
\end{gather}
Note that $\psi_i$ has the dimension $1/\sqrt{\text{Volume}}$ and hence $\sigma$ has the dimension $\sqrt{\text{Volume}}$.
The quantization of the free phonon field is more complicated due to the nonlinear nature of the interaction term (\ref{3.3}). In order to do the quantization, we look at the harmonic version of (\ref{3.3}), \ie we use Eq.(\ref{3.19}) up to order $\tilde\rho^2$.

We next find $H_0$ (Eq.(\ref{3.2})) to lowest order \ie up to second order in density. From Eqs.(\ref{3.20}), (\ref{04.6}) and (\ref{04.7}), we can easily obtain
\begin{equation}\label{04.8}
  \Phi = \sum i\frac{v_{s}^2}{f_i} \psi_i(\vec r)\mathrm{e}^{-if_it} \sigma_i + c.c.,
\end{equation}
and hence for a given mode
\begin{align}\label{04.9}
  \int\mathcal{H}_0\md^3r &= -\frac12\rho_0 \int \Phi\nabla^2\Phi\md^3r \nonumber \\
  &= \frac{f^2}{2v_s^2}\rho_0 \int\Phi^2\md^3r = \rho_0v_s^2|\sigma|^2.
\end{align}
We thus quantize the phonon field via
\begin{equation}\label{04.10}
  \sigma_i \to \sqrt{\frac{\hbar f_i}{2\rho_0v_s^2}}b_i,
\end{equation}
where $b_i$ is the Bosonic annihilation operator for the phonon with frequency $f_i$. The unperturbed Hamiltonian for the phonon field
\begin{equation}
  \tilde\rho = \sum \psi_i(\vec r)\sqrt{\frac{\hbar f_i}{2\rho_0v_s^2}}b_i \mathrm{e}^{-if_it} + c.c., \label{04.11}
\end{equation}
is
\begin{equation}
  H_0 = \sum_i\hbar f_i b^\dag_i b_i + \mathcal{O}(\tilde{\rho}^3). \label{04.111}
\end{equation}
The terms of the order $\tilde{\rho}^3$ can be obtained by using the expansion (\ref{3.19}). These correspond to three phonon scattering processes. The quantized form of the interaction Hamiltonian (\ref{3.15}) can now be obtained by using Eqs.(\ref{04.4}) and (\ref{04.11}). The final result is dependent on the different modes involved in the optomechanical interactions and their overlap integrals. We write
\begin{equation}\label{04.12}
  H^\sem_1 = v_L + v_{NL},
\end{equation}
where $v_L$ and $v_{NL}$ are respectively the linear and nonlinear optomechanical interactions. In what follows, we drop all rapidly oscillating terms at twice the cavity frequencies.

The linear part then can be written as
\begin{align}
  v_L &= -\hbar \sum_{ijl}\left\{ g'_{ijl}\mathrm{e}^{i(\omega_i-\omega_j-f_l)t} a^\dag_i a_j b_l \right. \nonumber \\
  &\qquad \left. + g''_{ijl}\mathrm{e}^{i(\omega_i-\omega_j+f_l)t} a^\dag_i a_j b^\dag_l \right\}, \label{04.13}\\
  g'_{ijl} &= \sqrt{\frac{\hbar f_l\omega_i\omega_j}{8\rho_0v_s^2}}g_1 \int\psi_l(\vec r)[\vec u^{(i)*}(\vec r)\cdot \vec u^{(j)}(\vec r)]\md^3r,  \label{04.14}
\end{align}
and $g''$ is obtained from $g'$ by replacing $\psi(\vec r)$ by $\psi^*(\vec r)$. For simplicity, we choose $\psi$ to be real and then we can drop the distinction between $g'$ and $g''$. The quantity $g'_{iil}$ is the frequency shift of the cavity mode for one photon. We can get an approximate estimate of $g'_{ijl}$ by using $\psi(\vec r)=1/\sqrt{V}$,
\begin{equation}\label{04.15}
  g'_{ijl} \approx \sqrt{\frac{\hbar f_l}{8\rho_0v_s^2}} \frac{\omega g_1}{\sqrt{V}}.
\end{equation}
For the fiber cavity`\cite{Jacobs} taking the mode volumn about $V\sim10^{-14}$m$^3$, $\omega$ corresponding to $1\mu$m, $f\sim2\pi\times10$MHz, $\rho_0v_s^2\approx 8213380$J/m$^3$, we find $g'_{iil}/g_1\approx2\pi\times30$kHz and hence $g'_{iil}\approx2\pi\times1.8$kHz. Thus, linear optomechanical interaction is quite significant and is comparable to that obtained with mechanical elements~\cite{Aspelmeyer,Vitali}.

The nonlinear part $v_{NL}$ has several contributions. We do not write all the terms but make an estimate. A term corresponding to two-phonon absorption has the form
\begin{equation*}
  -\hbar \sum_{ijl_1l_2} p_{ijl_1l_2} \mathrm{e}^{i(\omega_i-\omega_j-f_{l_1}-f_{l_2})t} a^\dag_i a_j b_{l_1}b_{l_2},
\end{equation*}
where $p_{ijl_1l_2}$ has the form
\begin{align}\label{04.16}
  p_{ijl_1l_2} &= \sqrt{\frac{\hbar f_{l_1}\omega_i\omega_j}{8\rho_0v_s^2}} g_2 \sqrt{\frac{\hbar f_{l_2}}{2\rho_0v_s^2}} \nonumber \\
  &\qquad \times \int\md^3r \psi_{l_1}(\vec r)\psi_{l_2}(\vec r) \vec u^{(i)*}\cdot \vec u^{(j)}.
\end{align}
Let us take $\vec u^{(i)}$ and $\vec u^{(j)}$ to be the same mode, then
\begin{equation}\label{04.17}
  p_{ijl_1l_2} \approx \sqrt{\frac{\hbar f_{l_1}}{8\rho_0v_s^2V}} \sqrt{\frac{\hbar f_{l_2}}{2\rho_0v_s^2V}} \omega g_2.
\end{equation}
Thus compared to the first-order optomechanical coupling, the second order is smaller by a factor
\begin{equation*}
  \sqrt{\frac{\hbar f_l}{2\rho_0v_s^2V}}\left(\frac{g_2}{g_1}\right) \approx 1\times10^{-10}\left(\frac{g_2}{g_1}\right) \sim 5\times10^{-12},
\end{equation*}
for parameters that were used in the estimation of linear optomechanical coupling. The second-order contribution has been generally found to be unimportant in most mechanical systems. However, considerable progress has been reported in achieving higher second-order coupling by placing the mechanical element at the crossing of two modes~\cite{Harris,Vitali}. One possibility to enhance $p$ would be to use nanometric volume for He. In view of the smallness of $p_{ijl_1l_2}$ compared to $g_{ijl}$, we drop the contribution $v_{NL}$ and work with
\begin{equation}\label{04.18}
  v_L = -\hbar \sum_{ijl}\left\{g_{ijl}a^\dag_i a_j \mathrm{e}^{i(\omega_i-\omega_j)t} (b_l\mathrm{e}^{-if_l t} + b_l^\dag\mathrm{e}^{if_l t}) + h.c. \right\}.
\end{equation}

For two field modes and one acoustic mode and assuming that $\vec u^{(1)}=\vec u^{(2)}$, we can simplify Eq.(\ref{04.18}) to
\begin{align}\label{04.19}
  v_L &= -\hbar g\left( a^\dag_1a_1 + a^\dag_2a_2 + a^\dag_1a_2\mathrm{e}^{i(\omega_1-\omega_2)t} \right. \nonumber \\
   &\qquad \left. + a_1 a^\dag_2\mathrm{e}^{-i(\omega_1-\omega_2)t}\right) \left(b_l\mathrm{e}^{-if_l t} + b_l^\dag\mathrm{e}^{if_l t}\right).
\end{align}
This is the standard form of the optomechanical interaction. The terms like $(a^\dag_1a_2b\mathrm{e}^{i(\omega_1-\omega_2-f_l)t} + h.c.)$ describe the upconversion process where a phonon and a photon $\omega_2$ combine to produce a photon $\omega_1$ if $\omega_1>\omega_2$. The terms like $(a_1a^\dag_2b^\dag\mathrm{e}^{-i(\omega_1-\omega_2-f_l)t} + h.c.)$ describe a downconversion process where a photon $\omega_2$ and a phonon $f_l$ are produced form a photon of frequency $\omega_1$. The Hamiltonian (\ref{04.19}) also consists of nonresonant terms like $a_1^\dag a_1b$. These terms play a significant role in two-step two-phonon processes as discussed in the next section. Clearly, the previously discussed processes~\cite{OMS1,OMS2,OMS3,OMS4} in other optomechanical systems would also apply to optomechanics in superfluid He, since the linear coupling constant $g$ in Eq.(\ref{04.19}) is quite significant. The advantage of superfluid He is its very large coherence time of the phonon, which is especially useful in quantum processing applications like state transfer~\cite{stateconv} and quantum memories~\cite{EIT,storage,memory}.

\section{Optomechanical interactions in superfluid H\lowercase{e} involving two-step two-phonon processes}
We next consider the very interesting possibility of two-phonon absorption~\cite{NLR} in optomechanical interactions in superfluid He. This, in a sense, is the analog of two photon absorption in atomic systems. Let us consider the following two steps:
\begin{align*}
  &\text{phonon}(f_1) + \text{photon}(\omega_1) \to \text{intermediate photon}(\omega'), \\
  &\text{intermediate photon}(\omega') + \text{phonon}(f_2) \to \text{photon}(\omega_2).
\end{align*}
\begin{figure}[hpbt]
\includegraphics[width=0.3\textwidth]{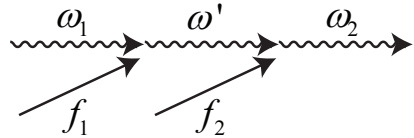}
\caption{ The two-step two-phonon anti-Stokes process.} \label{fig1}
\end{figure}
This two-step process shown in Fig.~\ref{fig1} is resonant if
\begin{equation}\label{6.1}
  \omega_1 +  f_1  +  f_2  = \omega_2,
\end{equation}
and is a combination of two upconversion processes. The two-step process is mediated by an intermediate photon of frequency $\omega'$. Such a two-step process can be significant as the intermediate photon can have the same frequency as the strong input photon $\omega_1$. Such a contribution comes form the term $a_1^\dag a_1b_1$ in the interaction (\ref{04.18}). For the present problem, the interaction $v_L$ can be written as
\begin{align}\label{6.2}
  v_L &= -\hbar g(a_1+a_2\mathrm{e}^{-i(\omega_2-\omega_1)t})^\dag (a_1+a_2\mathrm{e}^{-i(\omega_2-\omega_1)t}) \nonumber \\
  &\qquad (b_1\mathrm{e}^{-if_1t}+b_2\mathrm{e}^{-if_2t} + h.c.).
\end{align}
Let the initial state be $|n_1,n_2,\mu_1,\mu_2\rangle$ which has $n_1(n_2)$ photons of frequency $\omega_1(\omega_2)$ and $\mu_1(\mu_2)$ phonons of frequency $f_1(f_2)$. The final state is $|n_1-1,n_2+1,\mu_1-1,\mu_2-1\rangle$. The transition probability for this process can be obtained by using second order Fermi Golden rule
\begin{equation}\label{6.3}
  R^{[2]} = \frac{2\pi}{\hbar} \delta(E_f-E_i) \left| \sum_j\frac{\langle f|v_L|j\rangle\langle j|v_L|i\rangle}{E_j-E_i}\right|^2,
\end{equation}
where $|j\rangle$ is the allowed intermediate state. The four important intermediate states with the corresponding energies are
\begin{gather*}
  |n_1-1,n_2+1,\mu_1-1,\mu_2\rangle,\quad E_j-E_i = \hbar(\omega_2-\omega_1-f_1); \\
  |n_1,  n_2,  \mu_1-1,\mu_2\rangle,\quad E_j-E_i = -\hbar f_1; \\
  |n_1,  n_2,  \mu_1,\mu_2-1\rangle,\quad E_j-E_i = -\hbar f_2; \\
  |n_1-1,n_2+1,\mu_1,\mu_2-1\rangle,\quad E_j-E_i = \hbar(\omega_2-\omega_1-f_2).
\end{gather*}
Using these intermediate states and Eq.(\ref{6.2}), the two-phonon absorption rate is calculated to be
\begin{align}\label{6.4}
  & \quad R^{[2]} =  \nonumber \\
  & \frac{2\pi g^4(2\gamma/\pi)}{[(2\gamma)^2+(\omega_1-\omega_2+f_1+f_2)^2]} \mu_1\mu_2n_1(n_2+1)(n_1+n_2)^2 \nonumber \\
  &\quad \cdot \left| \frac{1}{(-\omega_1+\omega_2-f_1)} - \frac{1}{f_1} - \frac{1}{f_2} + \frac{1}{(-\omega_1+\omega_2-f_2)} \right|^2,
\end{align}
where we introduced the width $\gamma$ for the phonon distribution. Note that the sum over intermediate states vanishes. Thus there is interference between different quantum pathways. Such interference effects are well-known in atomic physics in the context of two photon processes (see Sec.7.5 in \cite{OMS1}). It is also known that relaxation effects generally make perfect interference imperfect leading to nonzero transition amplitudes. For our system in a cavity, the cavity line width $\kappa$ is an important factor and it makes $R^{[2]}$ nonzero. The denominators like $1/f$ need to be modified by inclusion of the phonon line width $\gamma$ which is much smaller than $\kappa$. The denominators depending on $\omega_1$ and $\omega_2$ get modified by inclusion of $\kappa$. A simple argument then modifies Eq.~(\ref{6.4}) to
\begin{align}\label{6.5}
  R^{[2]} &= \frac{2\pi g^4(2\gamma/\pi)}{(2\gamma)^2+(\omega_1-\omega_2+f_1+f_2)^2} \nonumber \\
  &\qquad \cdot \left(\frac{\kappa}{f}\right)^2 \frac{1}{f} \mu_1\mu_2n_1(n_2+1)(n_1+n_2)^2,
\end{align}
This should be compared with the corresponding result $R^{[1]}$ for one phonon absorption
\begin{equation}\label{6.6}
  R^{[1]} = \frac{2\pi g^2(\gamma/\pi)}{\gamma^2+(\omega_1-\omega_2+f_1)^2} \cdot \mu_1n_1(n_2+1),
\end{equation}
which is easily obtained from Eq.(\ref{04.18}). Let us compare the strength of $R^{[2]}$ with $R^{[1]}$ at resonance
\begin{align}\label{6.7}
  \frac{R^{[2]}}{R^{[1]}} &\approx \frac{g^2}{2} (\frac{\kappa^2}{f^4}) \cdot (n_1+n_2)^2\mu_2\nonumber \\
  &\sim  \frac{g^2\kappa^2}{2f^4} n_1^2\mu_2 \qquad \text{as } n_1\gg n_2.
\end{align}
For $g\sim2\pi\times20$Hz, $f\sim2\pi\times10$MHz, $\kappa/f\sim1/10$,
\begin{align}\label{6.8}
  \frac{R^{[2]}}{R^{[1]}} &\approx 2\times10^{-16} n_1^2\mu \nonumber \\
  &= 2\times10^{-4} \mu  \qquad \text{for } n_1\sim10^{6}.
\end{align}
For temperatures of the order of $10$mK, $\mu_2\approx10$ and hence $R^{[2]}/R^{[1]}\sim10^{-3}$, leading to substantial probability for two-step two-phonon absorption. Note that, instead of using thermal phonons, we can inject phonons from an external source~\cite{Wright}.

\section{Conclusions}
In conclusion, we have developed a first principle theory of the optomechanical interactions in superfluid He. The theory is formulated in terms of the superfluid density field, so that multimode phonon optomechanics can be studied. The intrinsic nonlinearities of superfluid are included.  We presented estimates of the strength of the optomechanical interactions and derived the canonical form of the Hamiltonian for linear optomechanical interactions. Using such canonical Hamiltonian standard effects like normal mode splitting, electromagnetically induced transparency in superfluid optomechanics can be studied. We also showed the importance of the two-step two-phonon process in superfluid He. The superfluid also has the possibility of nonlinear phonon processes which one can integrate with the optomechanical processes. For example, two phonons $f_1$ and $f_2$ can combine via the cubic nonlinearity in Eq.(\ref{3.19}) and the generated phonon can be used for optomechanical interactions.

\section*{Acknowledgment}
S.S.J. acknowledges the hospitality of the Oklahoma State University, while this work was done. The authors thank Kenan Qu for his support in preparing the paper. G.S.A. thanks J. Harris for preliminary correspondence.


\begin{thebibliography}{99}
\bibitem{PNAS} N. Kiesel, F. Blaser, U. Deli\'{c}, D. Grass, R. Kaltenbaek, and M. Aspelmeyer, PNAS, \textbf{110}, 14180 (2013).
\bibitem{Barker} T. S. Monteiro, J. Millen, G. A. T. Pender, F. Marquardt, D. Chang, P. F. Barker, New J. Phys. \textbf{15}, 015001 (2013).
\bibitem{Aspelmeyer} S. Groeblacher, K. Hammerer, M. R. Vanner, M. Aspelmeyer, Nature(London) \textbf{460}, 724 (2009).
\bibitem{Vitali} M. Karuza, M. Galassi, C. Biancofiore, C. Molinelli, R. Natali, P. Tombesi, G. Di Giuseppe, and D. Vitali, J. Opt. {\bf 115}, 025704 (2013).
\bibitem{Harris} D. Lee, M. Underwood, D. Mason, A. B. Shkarin, S. W. Hoch, and J. G. E. Harris, arXiv:1401.2968 (2014).


\bibitem{EIT} G. S. Agarwal and S. Huang, Phys. Rev. A \textbf{81}, 041803(R) (2010); S. Weis, el al., Science \textbf{330}, 1520 (2010); J. D. Teufel, el al., Nature (London), \textbf{471}, 204 (2011); Y. Liu, M. Davan\c{c}o, V. Aksyuk, and K. Srinivasan, Phys. Rev. Lett. \textbf{110}, 223603  (2013); Kenan Qu and G. S. Agarwal, Phys. Rev. A \textbf{87}, 031802 (2013); A. H. Safavi-Naeini, \etal, Nature (London), \textbf{472}, 69 (2011).
\bibitem{storage} V. Fiore, Y. Yang, M. C. Kuzyk, R. Barbour, L. Tian, and H. Wang, Phys. Rev. Lett. {\bf 107}, 133601 (2011); C. Dong, V. Fiore, M. C. Kuzyk, and H. Wang, Phys. Rev. A {\bf 87}, 055802 (2013); V. Fiore, C. Dong, M. C. Kuzyk, and H. Wang, ibid {\bf 87}, 023812 (2013).

\bibitem{Schwab} L. A. DeLorenzo, and K. C. Schwab, arXiv:1308.2164 (2013).
\bibitem{meeting} N. E. Flowers-Jacobs, A. D. Kashkanova, A. B. Shkarin, S. W. Hoch, C. Deutsch, J. Reichel  and J. G. E. Harris, in ``Meeting of The American Physical Society'', (Denver, Colorado, BAPS.2014.MAR.Q35.4).
\bibitem{Jacobs} N. E. Flowers-Jacobs, S. W. Hoch, J. C. Sankey, A. Kashkanova, A. M. Jayich, C. Deutsch, J. Reichel, J. G. E. Harris, Appl. Phys. Lett. \textbf{101}, 221109 (2012).

\bibitem{OMS1} G. S. Agarwal, Quantum Optics (Cambridge University Press, 2012), Chap. 20.
\bibitem{OMS2} M. Aspelmeyer, T. J. Kippenberg, and F. Marquardt, arXiv:1303.0733 (2013).
\bibitem{OMS3} C. Genes, A. Mari, D. Vitali, and P. Tombesi, Adv. At., Mol. Opt. Phys. \textbf{57}, 33 (2009).
\bibitem{OMS4} M. Aspelmeyer, P. Meystre and K. Schwab, Phys. Today \textbf{65}, 29 (2012).

\bibitem{London} F. London, Superfluids (Wiley, New York, 1950).
\bibitem{Wright} D. R. Wright, J. S. Foster, B. Hadimioglu, and C. F. Quate, J. Appl. Phys. \textbf{68}, 4438 (1990).
\bibitem{Abraham} B. M. Abraham, Y. Eckstein, J. B. Ketterson, M. Kuchnir, and P. R. Roach, Phys. Rev. A \textbf{1}, 250 (1970).
\bibitem{Stratton} J. A. Stratton, Electromagnetic Theory (McGraw Hill, Newyork, 1941), Eq.(35) on p.144.
\bibitem{Goldstein} H. Goldstein, C. P. Poole, and J. L. Safko, Classical Mechanics, 3rd Ed, (Addison-Wesley, Newyork, 2002), Sec. 13.4.
\bibitem{Meyer} C. Boghosian and H. Meyer, Phys. Rev. {\bf 152}, 200 (1966); R. J. Donnelly and C. F. Barenghi, J. Phys. Chem. Ref. Data, {\bf 27}, 1217 (1998).

\bibitem{stateconv} E. Verhagen, S. Del\'{e}glise, S. Weis, A. Schliesser, and T. J. Kippenberg, Nature(London) {\bf 482}, 63 (2012);
    L. Tian, Phys. Rev. Lett. {\bf 108}, 153604 (2012); Y. -D. Wang and A. A. Clerk, ibid {\bf 108}, 153603 (2012);
    T. A. Palomaki, J. W. Harlow, J. D. Teufel, R. W. Simmonds, and K. W. Lehnert, Nature {\bf 495}, 210 (2013);
    R. W. Andrews, R. W. Peterson, T. P. Purdy, K. Cicak, R. W. Simmonds, C. A. Regal, and K. W. Lehnert, Nature Phys. {\bf 10}, 321 (2014).
\bibitem{memory} A. I. Lvovsky, B. C. Sanders and W. Tittel, Nat. Photon. \textbf{3}, 706 (2009).


\bibitem{NLR} Sumei Huang and G. S. Agarwal, Phys. Rev. A {\bf 83}, 023823 (2011); Y. -C. Liu, Y. -F. Xiao, Y. -L. Chen, X. -C. Yu, and Q. Gong, Phys. Rev. Lett. \textbf{111}, 083601 (2013);
    A. Kronwald and F. Marquardt, ibid. \textbf{111}, 133601 (2013);
    M. -A. Lemonde, N. Didier, and A. A. Clerk, ibid. \textbf{111}, 053602 (2013);
    K. B{\o}rkje, A. Nunnenkamp, J. D. Teufel, and S. M. Girvin, ibid. \textbf{111}, 053603 (2013).


\end{thebibliography}
\end{document}